\documentstyle[12pt,prc,aps,psfig]{revtex}
\tighten
\begin{document}
\draft

\title{Comment on                                        \\
``Unmasking the nuclear equation of state''}
\author{F. Sammarruca}
\address{Physics Department, University of Idaho, Moscow, ID 83844, U.S.A}
\date{\today}
\maketitle
\begin{abstract}
Referring to a recently published paper [J. Piekarewicz, Phys. Rev. C 69, 041301 (2004)],
we point out that Dirac-Brueckner-Hartree-Fock 
calculations of the symmetry energy predict values of the neutron skin 
of heavy nuclei which are remarkably consistent with those from 
a recent ``best-fit'' analysis. 
\end{abstract}
\narrowtext
\vspace*{0.5cm}

This is a brief comment on the paper ``Unmasking the nuclear equation of
state'' by J. Piekarewicz \cite{Piek04}.

In that article, ``accurately calibrated'' relativistic mean field models
are used to calculate the distribution of isoscalar-monopole strength in 
$^{90}$Zr and
$^{208}$Pb as well as the isovector-dipole strength in
$^{208}$Pb using a continuum RPA approach.                   
One of the main results of that analysis is that the neutron skin of 
$^{208}$Pb is constrained to the range $S_n \leq 0.22$ fm. More precisely, the 
suggested values for the skin of 
$^{208}$Pb and the (related) compression modulus of symmetric nuclear matter are         
 $S_n =0.17$ fm and $K = 248 \pm 8$ MeV, respectively.

In reviewing the present status of the field, the author cites non-relativistic models
\cite{NR1,NR2,NR3} and relativistic ones \cite{MF1,MF2,MF3}, 
 and points out that, consistently, smaller/larger values of $K$ and
the neutron skin are obtained in the former/latter case.
In fact, one of the author's conclusions is that the results of his analysis           
seem to favor somewhat the non-relativistic predictions. 

We would like to draw attention to the fact that there 
are approaches beyond nonrelativistic 
(such as Skyrme) models and relativistic mean-field models (``calibrated'' or not).
Both schemes discussed in Ref.~\cite{Piek04}, relativistic and nonrelativistic,
use effective forces. On the other hand, the approach of Ref.~\cite{AS02} is {\it microscopic}    
(as it starts from realistic free-space two-body forces), {\it and 
 relativistic}  
(as it takes properly into account 
the Dirac structure of the nucleon).                                            
In Ref.~\cite{AS203}, we apply the 
equation of state for isospin-asymmetric matter described in Ref.~\cite{AS02}
to predict neutron radii and neutron skins. The values for the compression modulus
of symmetric matter and the skin of 
$^{208}$Pb that we predict with Bonn-B are $233$ MeV and $0.188$ fm, respectively. These  
are obtained from an equation of state that contains {\it no
free parameters}. We also notice that they are remarkably                           
close to the values suggested by Piekarewicz as most realistic.    
We think this point deserves interest.         

Thus, although we obviously agree with the 
quantitative conclusions of Ref.~\cite{Piek04}, we disagree with the    
implications concerning what a desirable theoretical approach to (symmetric or
asymmetric) nuclear matter should be.

\end{document}